# Investigation of the electrostatic potential of a grain boundary in Y-substituted BaZrO$_3$ using inline electron holography

Tarjei Bondevik[1], Heine H. Ness[1], Calliope Bazioti[1], Truls Norby[1], Ole Martin Løvvik[1,2], Christoph T. Koch[3], Øystein Prytz[1]

[1]Centre for Materials Science and Nanotechnology, University of Oslo, Norway

[2]SINTEF Materials Physics, Oslo, Norway

[3]Department of Physics, Humboldt-Universität, Berlin, Germany

## Abstract

We apply inline electron holography to investigate the electrostatic potential across an individual BaZr$_{0.9}$Y$_{0.1}$O$_3$ grain boundary. With holography, we measure a grain boundary potential of -1.3 V. Electron energy loss spectroscopy analyses indicate that barium vacancies at the grain boundary are the main contributors to the potential well in this sample. Furthermore, geometric phase analysis and density functional theory calculations suggest that reduced atomic density at the grain boundary also contributes to the experimentally measured potential well.

## Introduction

With good chemical stability and high bulk proton conductivity, yttrium-doped barium zirconate (BZY) is the basis for state-of-the-art electrolytes for ceramic proton conducting fuel cells [1]. A major drawback is the low grain boundary (GB) conductivity, typically orders of magnitude lower than the bulk conductivity [2-6]. This is usually explained by strongly proton-depleted and hence highly resistive space charge layers. These are induced by positively charged defects segregating to the GB core to lower lattice mismatch strain there [4], generating a net electrostatic potential. The potential across the GB is hence a relevant parameter related to the low GB conductivity in BZY. Since the area of interest around the GB is in the range of only a few nanometers, direct experimental observation of the GB potential is challenging.

The low GB conductivity is generally revealed by impedance spectroscopy of polycrystalline samples which then represents average properties of a number of differently oriented GBs [4, 6, 7]. By application of several assumptions and knowledge of the grain size, the average effective thickness of GBs can be deduced, and from this the specific GB conductivity can be obtained. The GB potential giving rise to the charge carrier depletion can then be estimated from the ratio between the GB and bulk conductivity [8] over a wide temperature range with a relatively simple experimental setup.

However, individual variations in the potentials of the GBs will generally be hidden in impedance spectroscopy. To circumvent this, one might have used bi-crystals of BZY, but these have proven challenging to make, and instead, measurements with microelectrodes have been performed, extracting impedances from individual GBs [7]. In these experiments, however, there arise problems of contact resistance of the microelectrodes as well as possible current detours across other grains. As a result, it is difficult to know if one is measuring solely the electrical response across an individual GB.

Computationally, individual BZY grain boundaries have been studied with density functional theory (DFT), revealing a strong tendency of defect segregation towards the GB, and a corresponding potential barrier across the GB calculated after numerical equilibration of the GB region with the segregation energies as input [9-12]. Different types of GBs exhibit similar potential barriers, increasing the confidence in the results. However, most BZY grain boundaries have random orientations [2] whose description is incompatible with the DFT requirement of a small, periodic supercell. Hence, DFT calculations only tell part of the story.

Inline electron holography is a transmission electron microscopy (TEM) technique where the projected electrostatic potential across an individual GB can be directly measured with sub-nanometer resolution. When TEM electrons travel through a specimen they experience a phase shift related to the thickness and the potential of the specimen. While this phase information collapses when acquiring a conventional TEM image, it is recovered in inline holography by using several defocused images. The technique has previously recovered a negative potential in the core of GBs in other polycrystalline materials [13-15].

Although measuring the electrostatic potential with TEM resolution is attractive, inline holography also has several challenges. First, the sample preparation and experimental setup is considerably more intricate than an impedance spectroscopy experiment or a DFT calculation. Second, the technique relies on a complex reconstruction algorithm to recover the collapsed phase information, necessitating careful post-analysis of the experiment. Third, inline holography is sensitive to all contributions to the electrostatic potential, including those less relevant for defect concentrations in the space charge layers. In addition, sample bending, or changes in specimen thickness may also affect the phase and amplitude of the transmitted electron beam. This requires a careful interpretation of the results. Despite its complications, holography gives insight in GB potentials different from what impedance and DFT studies can provide.

In this work, we apply inline electron holography to measure the electrostatic potential across an individual BZY grain boundary. We discuss how segregation of defects and structural distortions affect the measured GB potential, using electron energy loss spectroscopy (EELS), energy dispersive X-ray spectroscopy (EDX), geometric phase analyses (GPA) and DFT calculations to complement the analysis. Finally, we discuss the results' implications for the electrical properties of the GB, and the utility of holography for understanding such properties.

For a clear discussion, we define three terms referring to different parts of the grain boundary, used extensively in this work. The *interface* of the GB refers to the infinitely thin (two-dimensional) boundary between two grains, the *core* refers to the area around the GB interface that is structurally distorted relative to the bulk structure, and the GB *region* comprises of the GB core and space charge layers in the bulk-like structure outside of the GB core.

# Theory

## The electrostatic potential of a crystal

At any position in a material, the total electrostatic potential $V_{\text{tot}}$ can be divided into four contributions:

$$V_{\text{tot}} = V_{\text{MIP}} + V_{\text{E}} + V_{\text{XC}} + V_{\text{fields}}. \tag{1}$$

Here, $V_{\text{MIP}}$ is the mean inner potential, $V_{\text{E}}$ the potential due to redistribution of free charge carriers, $V_{\text{XC}}$ the exchange correlation potential, and $V_{\text{fields}}$ the potential caused by electrostatic fields in and around the specimen. The specimen's charge distribution is largely unaffected by the high energy TEM electrons (hereafter: fast electrons) passing through it, approximating $V_{\text{XC}}$ measured by TEM to zero [16]. Furthermore, we assume that any changes in $V_{\text{fields}}$ from the microscope itself or due to charging of the specimen during beam illumination are constant across a GB region of roughly 10 nm. This leaves local variations in $V_{\text{MIP}}$ and $V_{\text{E}}$ as the relevant quantities to consider in this work.

The mean inner potential of a finite crystal is generally positive relative to vacuum, and, within the independent atom approximation, can be expressed as [17]

$$V_{\text{MIP}} = \frac{h^2}{2\pi m_0 e \Omega} \sum_j n_j f_{\text{el}}^j(0), \tag{2}$$

where the summation includes all atomic species within the material, $h$ is Planck's constant, $m_0$ the rest mass of the electron, $e$ the elementary charge, $\Omega$ the unit cell volume, $n_j$ the number of atoms of species $j$ per unit cell, and $f_{\text{el}}^j(0)$ the electron scattering factor at zero scattering angle for atoms of species $j$. Equation (2) reveals at least three processes that may cause variations in the local mean inner potential at a GB interface, shown qualitatively in Figure 1. Part (a) shows how vacancies at the GB will lower the number of atoms per unit cell $n_j$ and hence lower $V_{\text{MIP}}$. In part (b), reduced atomic density due to lattice distortions at the GB increases the unit cell volume $\Omega$ locally and reduces $V_{\text{MIP}}$. Part (c) shows a substitutional foreign atom in the GB core with a different electronic scattering factor, hence yielding a different $V_{\text{MIP}}$. Also, a local rearrangement of charges due to bonding may affect $V_{\text{MIP}}$, however, in most cases to a lesser degree than the former three effects.

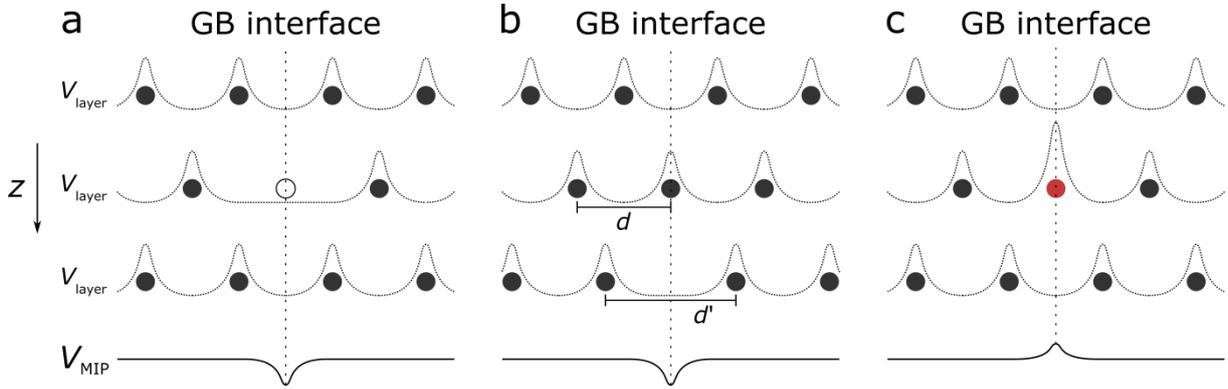

Figure 1. Qualitative description of three main types of phenomena that can affect the mean inner potential at the GB: segregation of vacancies (a), lattice distortions with $d' > d$ (b), and segregation of substitutional foreign atoms(c). The black, white and red disks represent host atoms, vacancies, and substituted atoms, respectively. The mean inner potential ($V_{\text{MIP}}$) is the average of the potentials at each atomic layer ($V_{\text{layer}}$), and needs to be summed for all atoms in the compound (Ba, Zr, Y, O) according to equation (2).

The other major contribution to the total potential, $V_{\text{E}}$, originates in a redistribution of free charge carriers. For instance, a positively charged GB core will induce a redistribution of free charge carriers in its vicinity, resulting in negatively charged space charge layers. Due to the large bandgap of BZY, protons and oxygen vacancies are the predominant free charge carriers. The relation between the concentrations of the free charge carriers and $V_{\text{E}}$ can be calculated with Poisson's equation,

$$\nabla^2 V_{\text{E}} = -\frac{\rho}{\epsilon}, \qquad (3)$$

where $\rho$ is the charge density and $\epsilon$ the dielectric constant. A procedure for numerically solving Poisson's equation to obtain $V_{\text{E}}$ in the space charge layer is given in Appendix B. Net positive charge at the GB core yields a positive $V_E$ across the GB, and vice versa for a negatively charged core. The positive GB core potential obtained for BZY with impedance spectroscopy in the literature stems from redistribution of free charge carriers and is hence represented by $V_{\text{E}}$, not $V_{\text{tot}}$, in equation (1).

In a holography experiment it is the change in total electrostatic potential $V_{\text{tot}}$ relative to some reference position that is measured. Separating its main contributions given in equation (1) is challenging when interpreting holography results. For instance, oxygen vacancies with an effective positive charge have two opposite effects on the total potential: $V_{\text{MIP}}$ will decrease because of the vacancies (ref. Figure 1a)) while $V_E$ increases because of the vacancies' effective positive charge, making it non-trivial to find their net contribution to the total potential.

To quantify the GB potential, the bulk potential is chosen as the reference such that

$$\Delta V_{\text{tot}}(x) \equiv V_{\text{tot}}(x) - V_{\text{tot,bulk}}, \quad (4)$$

and similarly for $V_E$ and $V_{\text{MIP}}$. We report the potentials in two distinct ways. Most commonly, the magnitude of the GB potential is reported. However, the spatial resolution of the DFT calculated potential is 0.01 Å, around three orders of magnitude finer than the experimental resolution. Hence, the DFT calculated potential will fluctuate strongly around the atomic nuclei, and comparing the magnitude of its GB potential with more smeared out experimental data may not be meaningful. Alternatively, comparing with the 2-dimensional projected potential defined as

$$\Delta V_{\text{tot}}^{2D} \equiv \int_{-\infty}^{\infty} \Delta V_{\text{tot}}(x) \, dx, \quad (5)$$

where $x$ is the direction perpendicular to the GB plane, may be more fruitful. The dimension of $V_{\text{tot}}^{2D}$ is voltage times length; we will report this in units of V·nm.

### Inline holography

The fast electrons travelling through the specimen are accelerated by the electrostatic potential of the specimen, giving rise to a phase shift $\Delta\phi(x,y)$. Within the phase object approximation and assuming a non-magnetic specimen, this phase shift is

$$\Delta\phi(x,y) = C(E) \int_{\text{path}} V_{\text{tot}}(x,y,z) dz \approx C(E) \cdot t(x,y) \cdot V_{\text{tot}}(x,y), \quad (6)$$

where $z$ is the optical axis, $V_{\text{tot}}(x,y)$ the average of the potential $V_{\text{tot}}(x,y,z)$ through the specimen, $t(x,y)$ the sample thickness, and $E$ the electron beam energy. The electron interaction constant is

$$C(E) = \frac{2\pi}{\lambda} \frac{m_0 c^2 + E}{E(2m_0 c^2 + E)}, \quad (7)$$

where $\lambda$ is the electron wavelength. As seen in equation (6), the electrostatic potential $V_{\text{tot}}(x,y)$ is obtained if we can find the local specimen phase shift and thickness. The phase shift can be found by using the Full-Resolution Wave Reconstruction (FRWR) software, briefly described in the Methodology section of this paper. The thickness is obtained by

$$t(x,y) = \lambda_{\text{mean}}^{\text{BZY}} \ln\left[\frac{I(x,y)}{I_0(x,y)}\right] \quad (8)$$

where $\lambda_{\text{mean}}^{\text{BZY}}$ is the inelastic mean free path of electrons going through the specimen, $I(x,y)$ is the intensity of an unfiltered image, and $I_0(x,y)$ is the intensity of the zero-loss filtered image. The inelastic mean free path of BZY was estimated to be $\lambda_{\text{mean}}^{\text{BZY}} = 123$ nm using the algorithm given in [18].

## Methodology

### Experimental details

The TEM specimen was made from a piece of the BaZr$_{0.9}$Y$_{0.1}$O$_3$ sample sintered at 2200 °C by Duval et al. [19]. Electron transparent wedge samples were prepared by mechanical grinding and polishing in tripod polisher (Allied MultiPrep). Final thinning was performed by Ar ion milling with a Fischione Model 1010, and plasma cleaning with a Fischione Model 1020 was applied before the TEM investigations.

Holography experiments were performed using a JEOL 2100F equipped with a ZnO/W(000) field emission gun and a Gatan imaging filter (GIF) and operated at 200 kV. An energy filtered focal series of a GB, consisting of 29 images, was automatically acquired on a US1000 FTPX camera using the Full Resolution Wave Reconstruction (FRWR) plugin for Digital Micrograph (DM) [20]. The specimen thickness in the area of interest was estimated to be 11 nm. To reduce the effect of undesired, additional phase shifts caused by electron diffraction within the crystal, an objective aperture was used to isolate the (000) reflection from other reflections, limiting the angular spread and the spatial frequencies to 14.05 mrad and 5.23 nm$^{-1}$, respectively. The images were acquired using a slit width of 10 eV centered on the zero-loss peak of the EELS spectrum, thereby limiting the contribution of inelastically scattered electrons. A corresponding reference focal series was also acquired of a vacuum region away from the sample, this was done to compensate for any changes in electron flux density caused by changes in the objective pre-field induced by defocusing the objective lens. The reconstruction was performed using the FRWR code with the gradient flipping method implemented. The reconstruction parameters were set to match the experimental setup described above. Dead pixels were removed using the remove X-rays function built in standard DM 1.85.

High resolution (S)TEM imaging, EDX, and EELS were conducted on an FEI Titan G2 60-300 equipped with a CEOS DCOR probe-corrector, Super-X EDX detectors and a Gatan GIF Quantum 965 EELS Spectrometer. Observations were performed at 300 kV with a probe convergence angle of 31 mrad, using High-Angle Annular Dark Field (HAADF) detectors, and the resulting spatial resolution was approximately 0.08 nm. The EELS spectra were acquired with an energy dispersion of 0.1 eV/channel and the energy resolution measured using the full width at half maximum (FWHM) of the zero-loss peak was 1.2 eV. GPA was performed on high resolution (S)TEM images in order to extract lattice strain maps, using the Gatan GMS software suite.

## Computational details

The DFT calculations were performed with the VASP code [21], employing the generalized gradient approximation (GGA-PBE) [22] and projector augmented wave (PAW) potentials [23] on the two stoichiometric $BaZrO_3$ GB supercells shown in Figure 2. Details about the supercells are provided in Table 1. We applied *k*-point densities according to the Monkhorst-Pack scheme: $2 \times 2 \times 1$ for the (111) GB, and $3 \times 3 \times 1$ for the (210) GB. Ionic relaxations were performed with the supercell being allowed to vary in size, and were considered to be converged when the residual forces were smaller than $0.03$ eV Å$^{-1}$. The plane wave cut-off energy was set to 600 eV, yielding a numerical precision better than 0.1 meV for relative total electronic energies. After relaxation we calculated the local electrostatic potential of the supercell, where the exchange-correlation part was excluded. This gives results comparable to the TEM data, because that contribution can be neglected for the fast electrons [16].

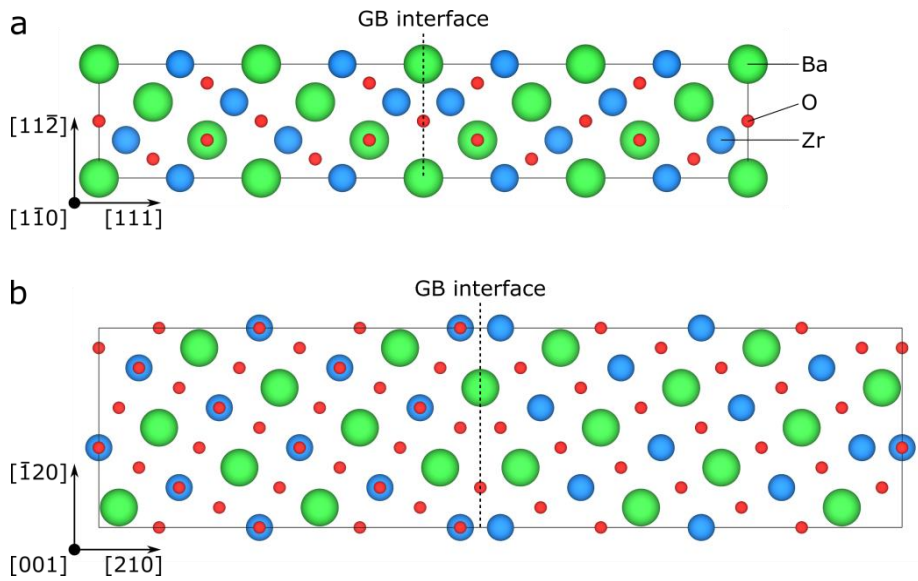

Figure 2. The $BaZrO_3$ GB supercells used in this study, with (a) showing the (111) GB used in [12], and (b) showing the (210) GB used in [11]. For clarity, the non-relaxed supercell is shown.

Table 1. Angles, dimensions, number of atoms, and GB separation for the two GBs studied with DFT, where $a_0$ is the lattice constant.

| GB | Angles | Dimensions | # of atoms | GB separation / nm |
|---|---|---|---|---|
| (111)[1$\bar{1}$0] | $\alpha = 90°, \beta = 90°, \gamma = 60°$ | $2\sqrt{2}a_0 \times 2\sqrt{2}a_0 \times 4\sqrt{3}a_0$ | 240 | 1.48 |
| (210)[001] | $\alpha = 90°, \beta = 90°, \gamma = 90°$ | $\sqrt{5}a_0 \times 3a_0 \times 4\sqrt{5}a_0$ | 300 | 1.95 |

# Results and discussion

## The measured electrostatic potential

Figure 3a shows a TEM image of a GB, with the area used for the FRWR reconstruction indicated by the red box. The specimen was tilted such that the electron beam was parallel to the GB interface. The reconstructed phase map is shown in Figure 3b, where a clear dip in phase was observed along the GB. By averaging 150 one pixel wide line profiles perpendicular to the GB and applying equation (6), we obtained the potential profile shown in Figure 3c. The GB potential had the magnitude $\Delta V_{\text{tot}}(0) = -1.3 \text{ V}$, with a 2-dimensional projected potential of $\Delta V_{\text{tot}}^{2D} = -2.6 \text{ V} \cdot \text{nm}$.

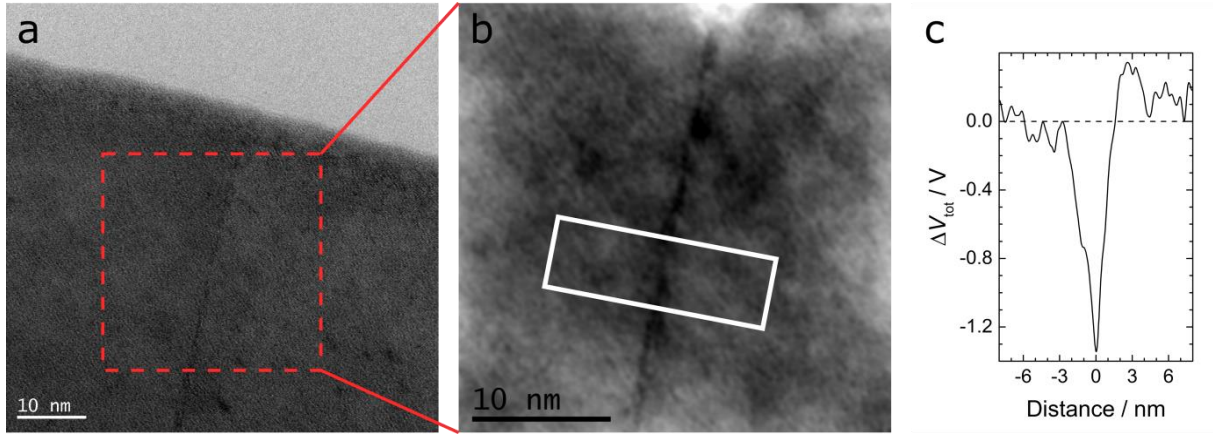

Figure 3. (a) TEM image of grain boundary. (b) Phase map, reconstructed from the red box in part (a). (c) Potential profile across the GB region, extracted from the 150 pixels wide white box in part (b). The grain boundary is randomly oriented, with Euler angles $\phi = 22.2°, \theta = 16.6°, \psi = 0.8°$ between the orientations of the two grains.

The literature based on impedance spectroscopy exclusively reports a positive GB potential for BZY [24]. This stems from the assumption that the GB core is positively charged with an excess of segregated oxygen vacancies and protons, and that the GB resistance is a result of the depletion of these charge carriers in the adjacent space charge layers.

At first glance, our holography result yielding a negative GB potential therefore seems to be in conflict with the established literature and model. However, when we recall that impedance spectroscopy reflects $\Delta V_E$ and holography reflects $\Delta V_{\text{tot}} \approx \Delta V_{\text{MIP}} + \Delta V_E$, and consider the special nature of this particular sample, we shall see that the negative GB potential we observe with holography is reasonable, and not in conflict with the existing model and literature.

Before discussing the origin of the potential well, a brief note will be made about charging of the specimen. BaZr$_{0.9}$Y$_{0.1}$O$_3$ is electronically insulating, and may hence charge during beam illumination, although no signs of this were observed in our high resolution TEM imaging. In the case of charging, the excess charge will induce an electrical field, meaning that $V_{\text{fields}} \neq 0$. We will, however, assume that any such effect is uniform across the length scale of interest (roughly 10 nm across the GB), and disregard $V_\text{fields}$ throughout this work.

### The origin of the potential well

To find the origin of the observed potential well as felt by the electron beam, we consider the local variations in $\Delta V_\text{MIP}$ and $\Delta V_\text{E}$ in the GB region. It is here fruitful to distinguish between two different phenomena affecting the GB potential. First, defects may segregate to the GB core, affecting both $\Delta V_\text{MIP}$ and $\Delta V_\text{E}$. Second, the GB may be free of point defects but have a lower mass density due to lattice distortions, affecting only $\Delta V_\text{MIP}$. In reality, a combination of these two effects may occur. We will first discuss the possibility of defects at the GB core, before considering possible lattice distortions later in this section.

#### *Defects at the grain boundary*

As seen in Figure 1a, any vacancy will reduce the local mean inner potential of the crystal. Also, any substitutional foreign atom will have a different electronic scattering factor $f_\text{el}^j(0)$ than the host atom and modify $V_\text{MIP}$, shown in Figure 1c. To determine how defects affect the potential, the relevant simple unassociated and fully ionized defects to consider are (in Kröger-Vink notation) the three vacancies ($v_\text{O}^{\bullet\bullet}$, $v_\text{Ba}''$, $v_\text{Zr}^{4'}$), protons in the form of protonated oxide ions ($\text{OH}_\text{O}^\bullet$), and yttrium doping on the zirconium site ($Y_\text{Zr}'$). Furthermore, loss of Ba during fabrication is believed to lead to occupation of Y on the Ba site ($Y_\text{Ba}^\bullet$) [25].

Figure 4a shows the barium core loss EELS signal (Ba M edge) across the GB region. An appreciable drop in the Ba signal is evident, and is not caused by any local thickness variation, as demonstrated by the $t/\lambda$ profile across the GB. We therefore conclude that the reduced Ba signal is caused by a lower concentration of Ba atoms in the GB region.

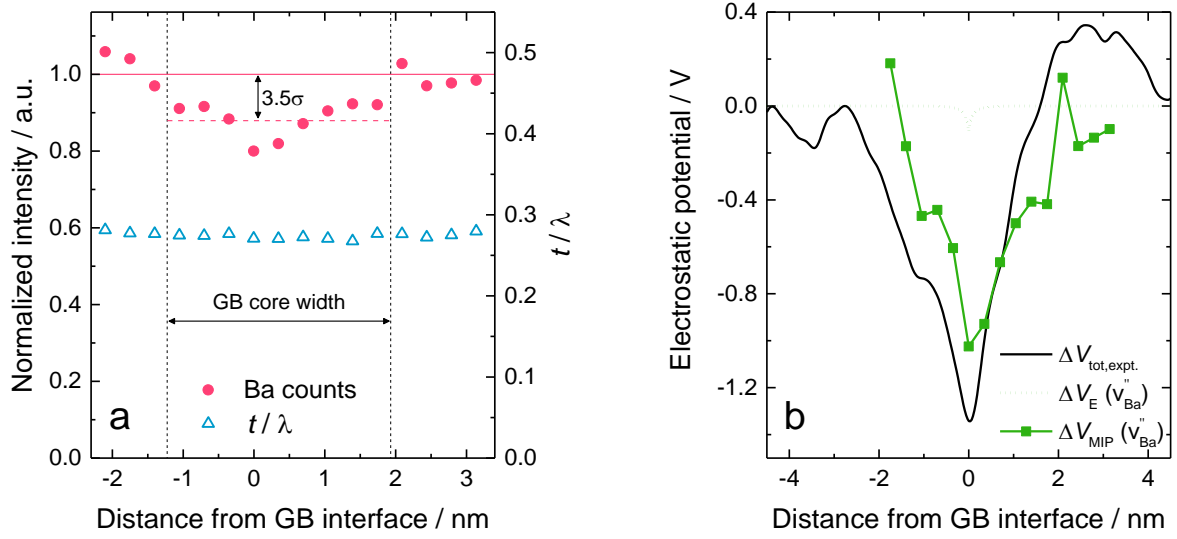

Figure 4. (a) Normalized barium core loss signal and TEM specimen thickness in the GB region, with $\sigma$ being the standard deviation of the Ba signal in the bulk region. The total number of counts per data point in the bulk region was roughly $3 \cdot 10^3$. The profile was acquired very close to the potential profile in Figure 3c, along the same GB. (b) Experimental holography potential $\Delta V_{\text{tot,expt.}}$, shown together with modelled $\Delta V_{\text{MIP}}$ and $\Delta V_{\text{E}}$, assuming the Ba core loss signal to be proportional to its concentration. To calculate $\Delta V_{\text{MIP}}$, the tabulated electronic scattering factor $f_{\text{el}}^{\text{Ba}^{2+}}(0) = 0.780$ nm was used [17]. To calculate $\Delta V_{\text{E}}$ we assumed $\epsilon_r = 75$ [9], $T = 300K$ and $[\text{OH}_O^{\bullet}] = 0$ in bulk (due to low $p\text{H}_2\text{O}$ in the TEM column: see Appendix A); a Jupyter Notebook of this calculation is provided at GitHub: https://github.uio.no/tarjeibo/bondevik-phd-thesis

If we assume that the barium signal is proportional to its concentration, and further that barium vacancies are the only segregating defect, we can calculate their effect on the potential. Modelling only one defect at a time does not give a realistic overall picture, but can demonstrate the effect of that particular defect. Using the Ba concentration profile in Figure 4a as input, $\Delta V_{\text{MIP}}$ and $\Delta V_{\text{E}}$ can be modelled from the equations (2) and (3), respectively, shown together with the experimental holography result in Figure 4b (details given in the caption). Note that the accuracy in $f_{\text{el}}^{\text{Ba}^{2+}}(0)$ used to model $\Delta V_{\text{MIP}}$ is limited, since isolated atoms were assumed and the redistribution of charges in the crystal environment was not considered.

From Figure 4b it is clear that the concentration of $v_{\text{Ba}}''$ has a much larger effect on $\Delta V_{\text{MIP}}$ than $\Delta V_{\text{E}}$, demonstrating how retrieving information about $\Delta V_{\text{E}}$ is challenging with electron holography. If the GB core is negatively charged from barium vacancies, the free charge carriers (that is, oxygen vacancies) will redistribute such that they contribute to the potential $\Delta V_{\text{E}}$ in Figure 4b. However, since electron holography measures $\Delta V_{\text{tot}} \approx \Delta V_{\text{MIP}} + \Delta V_{\text{E}}$, and in this case $|\Delta V_{\text{MIP}}| \gg |\Delta V_{\text{E}}|$, information about $\Delta V_{\text{E}}$ will be almost completely hidden.

Barium deficiency starting in the GBs tends to occur during sintering of BZY samples [25, 26], and should be expected in this sample as well, especially when considering the very high sintering temperature (2200 °C). Hence, the large Ba deficiency we observe at the GB core seems reasonable.

Several studies have reported yttrium segregation towards the GB core [4, 27, 28]. Figure 5a shows the yttrium concentration across the GB, measured with EDX. The results indicate an increased yttrium concentration at the GB core, either as $Y'_{Zr}$ segregating from bulk or $Y^{\bullet}_{Ba}$ due to loss of barium during fabrication. The concentration from Figure 5a is used to model both cases in Figure 5b. Again, the potential profiles are modelled based on the assumption that the defects (here: $Y'_{Zr}$ or $Y^{\bullet}_{Ba}$) are the only contributors to the total potential.

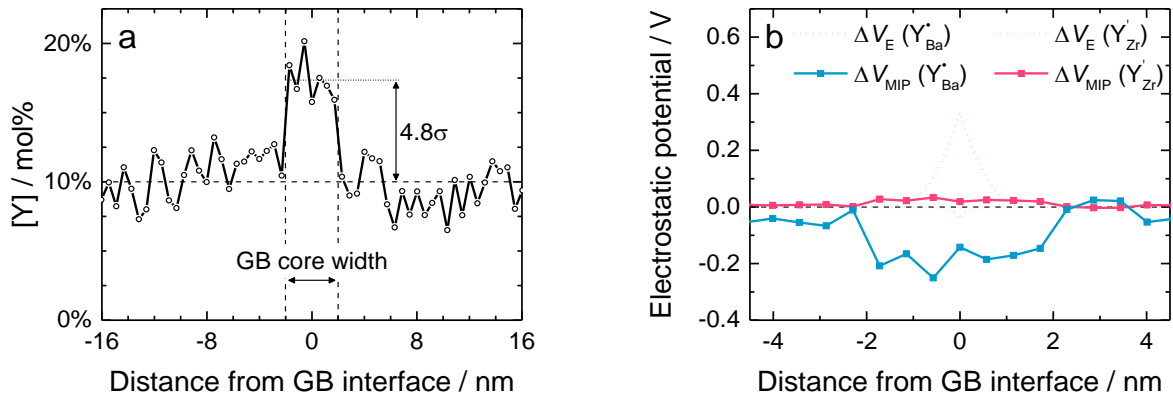

Figure 5. Part (a): EDX profile of the yttrium segregation, normalized to the known bulk concentration, with $\sigma$ being standard deviation of the bulk signal. The profile was acquired very close to the potential profile in Figure 3c, along the same GB. Part (b): Two independent modelling results of $\Delta V_{MIP}$ and $\Delta V_E$, assuming that $Y'_{Zr}$ is the system's only defect (red color), and $Y^{\bullet}_{Ba}$ is the system's only defect (blue color), modelled with the same method as in Figure 4b (using tabulated electronic scattering factors from [17] to model $\Delta V_{MIP}$).

Although considerable yttrium enrichment is observed in the GB core, its effect on the potential is modest. Segregation of $Y'_{Zr}$ has a negligible effect on both $\Delta V_{MIP}$ and $\Delta V_E$. $Y^{\bullet}_{Ba}$ enrichment in the GB core has a somewhat larger effect on both $\Delta V_{MIP}$ and $\Delta V_E$, contributing slightly to a negative 2-dimensional projected GB potential. Another important point is that part of the barium deficiency shown in Figure 4a may be in the form of $Y^{\bullet}_{Ba}$ rather than $v''_{Ba}$. If this is the case, the reported effect $v''_{Ba}$ has on $\Delta V_{MIP}$, shown in Figure 4b, is overestimated.

Figure 6a shows the zirconium core loss EELS signal in the GB region. No clear trends in the concentration profile are observed; however, the interpretation is challenging due to the high noise level. The oxygen signal, shown in Figure 6b, is less noisy, with no indications of any segregation of oxygen vacancies to the GB core. Finally, the bulk proton concentration can be calculated to be negligible in the high-vacuum TEM column (ref. Appendix A), and from that we assume that there was no significant proton accumulation in the GB core during our holography experiment.

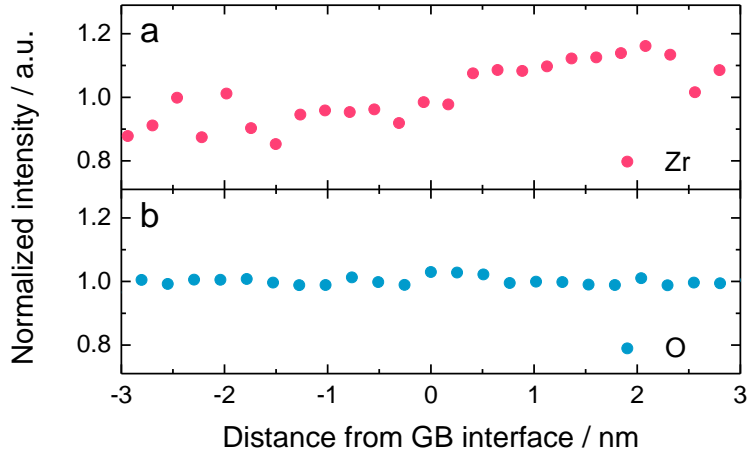

Figure 6. Normalized zirconium (a) and oxygen (b) core loss signal in the GB region, with the total number of counts being roughly $2 \cdot 10^4$ (a), and $6 \cdot 10^4$ (b). The profiles were acquired close to the potential profile in Figure 3c. Thickness measurements are not shown, but indicated constant thickness in the region. Principle component analysis was used on both datasets to reduce spectral noise.

To find which defects that contribute to the GB potential well, let us summarize our findings, shown in Table 2. There are no indications of neither $v_{Zr}^{4\prime}$ nor $v_O^{\bullet\bullet}$ accumulation at the GB core, although the EELS data has a high noise level for Zr. There may be considerable $Y_{Zr}'$ or $Y_{Ba}^{\bullet}$ accumulation, but they only have a small effect on $V_{tot}$. $OH_O^{\bullet}$ most likely has a negligible contribution due to the high vacuum inside the TEM column. This leaves $v_{Ba}''$ as the remaining defect that can explain the measured potential well.

Table 2. Potential wells and projected potentials. The holography potential is directly read from the graph in Figure 3c. The EELS and EDX potentials are found by summing contributions from $\Delta V_{MIP}$ and $\Delta V_E$ in Figure 4b and Figure 5b, respectively.

| Contribution to $\Delta V_{tot}$ | Technique | GB core width / nm | $\Delta V_{tot}(0)$ / V | $\Delta V_{tot}^{2D}$ / V·nm |
|---|---|---|---|---|
| All effects | Holography | 4.4* | -1.3 | -2.6 |
| $v_{Ba}''$ | EELS | 2.1† | -1.1 | -2.0 |
| $Y_{Zr}'$ | EDX | 2.8‡ | -0.01 | +0.08 |
| $Y_{Ba}^{\bullet}$ | EDX | 2.8‡ | +0.08 | -0.42 |
| $v_O^{\bullet\bullet}$ | EELS | N/A | 0 | 0 |
| $v_{Zr}^{4\prime}$ | EELS | N/A | N/A | N/A |

* Width of the region with $\Delta V_{tot} < 0$ from Figure 3c.

† Average width of the Ba deficient region in three EELS profiles (one of the profiles is shown in Figure 4a).
‡ Average width of the Y enriched region in two EDX profiles (one the profiles is shown in Figure 5a).

Given the similarity between the modelled $\Delta V_{\text{MIP}}$ and the experimental $\Delta V_{\text{tot}}$ in Figure 4b, barium vacancies are a plausible cause of the measured potential well. Further analyses suggest, however, that lattice distortions in the GB core may also contribute to the potential well.

### *Lattice distortion at the grain boundary*

Lattice distortions can be caused by at least two factors. First, local lattice distortions may occur at the GB. This is a simple geometrical argument: as two crystals join at the GB, the abrupt termination of each crystal leads to an energetically unfavorable structure. To reduce the energy of the crystal, atoms at the GB will relax into new positions, lowering the mass density. Second, repelling charged defects in the GB core may induce tensile strain, lowering the mass density and hence lowering the GB potential. Following the methodology laid out by Dunin-Borowski et al. [29], we calculated that tensile strain from repelling charged defects affects the potential with less than 0.01 V, and can thus be neglected. Effects from local lattice distortions on the other hand are significant.

To investigate the effect from local lattice distortions, we performed a GPA analysis. Such an analysis is challenging to perform across a GB since it requires the crystals to be in zone axis, which may be impossible to achieve simultaneously on both sides of the GB. Figure 7 shows a GPA analysis across the GB, where the grain on the right-hand side is in zone axis. The analysis suggests an increased lattice parameter in the direction perpendicular to the GB plane. The region of strain is about 1.3 nm wide into the grain on the right-hand side, leading to a 2-3 nm wide strained region if we assume similar strain on the other side of the GB interface. With strain resulting in lower atomic density, we expect a lowering of the mean inner potential. At the GB interface the GPA results claim an increased lattice parameter of 4 %, corresponding to $\Delta V_{\text{MIP}}(0) = -0.58$ V calculated from equation (2).

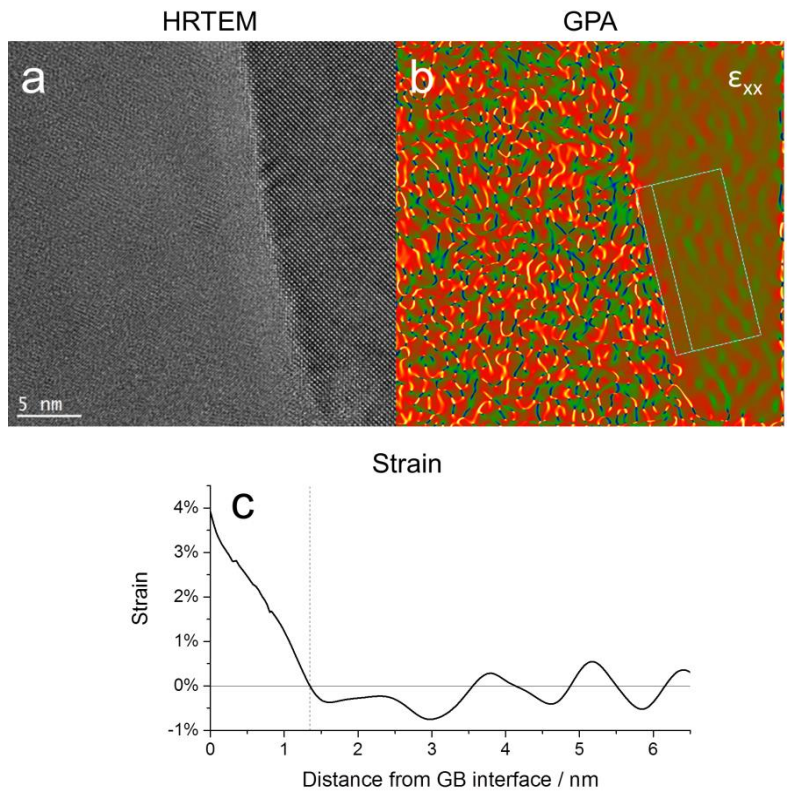

Figure 7. High resolution TEM image of the GB (a) with GPA strain map (b) and corresponding strain profile (c), showing the lattice expansion in the direction perpendicular to the GB, acquired from the indicated rectangle in part (b).

To complement the experimental data on lattice distortions, we used DFT to calculate the local potential on the stoichiometric (111) and (210) GBs, whose supercells are shown in Figure 2. These particular GBs were chosen because of their relatively small supercell sizes. Figure 8 shows the calculated potentials: each GB has a potential well in the GB core, explained by reduced mass density. In the (111) GB supercell, the mass density was 6.2 % lower in the indicated GB region compared to the bulk region from the same supercell. In the (210) GB supercell, the reduction in mass density in the GB region was as much as 13.2 %.

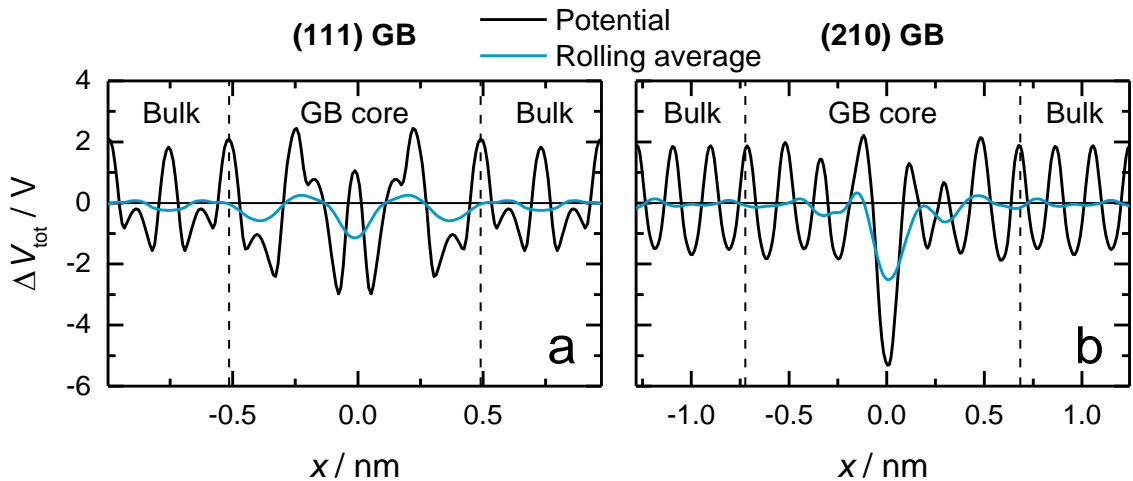

**Figure 8.** Potential calculated with DFT for the (111) GB (a), and the (210) GB (b), where the 3-dimensional potential $\Delta V_{\text{tot}}(x,y,z)$ has been projected to the *x*-axis. The GB core is defined as where the increase in interplanar distance (due to relaxation) between atomic planes parallel to the GB interface is more than 1%. As a guide for the eye a rolling average of the potential is shown, calculated as $\Delta V_{\text{rol.avg.}}(x) = \frac{1}{2d}\int_{x-d}^{x+d} \Delta V_{\text{tot}}(x)\mathrm{d}x$, where $d$ is the interplanar distance between the atomic planes parallel to the GB plane.

The DFT calculated potential wells (-3.0 and -5.3 V) were significantly deeper than the potential well obtained with holography (-1.3 V). This may be due to the simple GB models used in the DFT calculations, which may not be directly comparable to the complex, randomly oriented GB studied experimentally. Even if they did exhibit similar changes of the potential around the GB, experimental resolution limitations and imperfections may explain the discrepancy between modelling and experiment. The potential wells from the DFT calculations are only around 1 Å wide. If such narrow potential wells indeed exist in the real specimen, it is beyond the resolution limit of holography and will be smeared out in the experimental data. In addition, imperfect sample tilt such that the TEM electron beam is not entirely parallel with the GB plane may contribute to further smearing.

A point of confusion may be DFT results in literature exclusively showing a positive potential across the $BaZrO_3$ grain boundary [9, 11, 12, 30]. These results are not in conflict with our DFT simulations showing the opposite sign of the potential. The DFT work in literature considers $\Delta V_\text{E}$, calculated with space charge modelling where defect segregation energies are used as input data. To obtain the proton concentration in the space charge layer – which is a governing quantity of the GB conductivity – $\Delta V_\text{E}$ is the relevant potential to consider. Our simulations on the other hand consider $\Delta V_\text{tot}$ in a *defect free* supercell. They simply show how that the atomic density in the GB core is reduced during relaxation, and that the reduced density is linked to a negative $\Delta V_\text{tot}$ in the core. They do not, however, provide information about segregation and redistribution of defects. For example, the experimentally observed reduction in Ba occupancy not considered in these simulations is expected to further reduce the total GB potential.

Table 3 summarizes our findings, where both GPA analysis and DFT calculations indicate that reduced atomic density may contribute to the measured potential well. Note that the 2-dimensional projected potential obtained with holography has significantly larger magnitude, implying that lower mass density alone is unlikely to cause the entire potential well.

Table 3. Potential wells and projected potentials from experimental and theoretical work. The holography and DFT potentials are directly read from the graphs in Figure 3c and Figure 8. The GPA potential is found by assuming the strain in Figure 7c to be identical on both grains.

| Contribution to $\Delta V_\text{tot}$ | Technique | GB core width / nm | $\Delta V_\text{tot}(x=0)$ / V | $\Delta V_\text{tot}^\text{2D}$ / V · nm |
|---|---|---|---|---|
| All effects | Holography | 4.4[§] | -1.3 | -2.6 |
| All effects except defects | DFT, (111) GB | 1.0[**] | -3.0 | -0.26 |
| All effects except defects | DFT, (210) GB | 1.2[**] | -5.3 | -0.53 |
| Strain (lower mass density) | GPA | 2.7[††] | -0.58 | -0.73 |

[§] Width of the region with $V_\text{tot} < 0$ from Figure 3c.
[**] Width of the GB core region from relaxed supercells, defined in Figure 8.
[††] Width of the strained region from the GPA analysis in Figure 7c, assuming identical strain on both grains.

### The grain boundary resistance

Here, we discuss the results' implications for the electrical properties of our GB. The general view of BZY grain boundaries based on impedance spectroscopy and DFT calculations is that they have positively charged cores and adjacent resistive space charge layers. Our results suggest that our GB core may in fact be negative, such that this particular GB does not exhibit any resistive space charge layers. It is important to stress that the absence of space charge layer resistance only applies for this particular GB and this particular sample; other GBs in the sample or in other samples may be different.

We will, however, speculate that our results indicate reduced proton mobility in the GB core that may be significant for the GB resistance. All four experimental techniques (holography, EELS, EDX, GPA) show a GB core width of around 2 nm or more, resulting from either structural distortions or defects. This implies strongly reduced crystal symmetry in a 2 nm wide region. The relation between reduced symmetry and poor proton mobility is well established [1]. Computationally, nudged elastic band calculations show a reduction in GB proton mobility of 3-4 orders of magnitude at $T = 800$ K [30, 31]. Considering that roughly 20 proton jumps in a highly distorted crystal is needed for a proton to diffuse through a 2 nm wide core, the usual assumption of constant proton mobility through the GB region seems somewhat unlikely. Whether the reduced proton mobility has a significant contribution to the aggregated GB resistance such that the space charge model should be modified, is, however, beyond the scope of this work.

### The utility of holography to determine space charge

Finally, we discuss the utility of electron holography to describe space charge related phenomena in materials like BZY. The space charge model relies upon redistribution of free charge carriers, hence the relevant potential for describing space charge is $\Delta V_\text{E}$, not $\Delta V_\text{MIP}$. To find information about $\Delta V_\text{E}$ across a GB, holography is not necessarily an effective method. This is because $\Delta V_\text{MIP}$ in many cases dominates $\Delta V_\text{E}$, making it challenging to extract reliable information about $\Delta V_\text{E}$. One therefore has to be cautious about relating the potential measured with holography to space charge in the GB region. In cases of low defect densities, the space charge layer and thus $\Delta V_\text{E}$ may, however, be extended much further than the extent of $\Delta V_\text{MIP}$, providing some chance to still separate the two effects.

The high resistance normally resulting from the space charge model also relies upon positive carriers like oxygen vacancies and protons segregating to the GB core. A tempting conclusion is that our reported $\Delta V_\text{tot}(0) = -1.3$ V $< 0$ implies that protons should feel an electrostatic attraction to the GB core. Although this may be true for our particular GB, since $\Delta V_\text{tot}(0) < 0$ most likely is caused by accumulation of $\text{v}_\text{Ba}''$ that attract protons, it does not hold in general. For example, if the GB core rather is filled with $\text{Y}_\text{Ba}^\bullet$ and has low atomic density, it may simultaneously exhibit $\Delta V_\text{tot}(0) < 0$ and still be positively charged, repelling protons. Hence, a holography result showing a deep potential well in the GB core does not necessarily imply that protons will segregate to the core.

## Conclusions

1) A potential of −1.3 V from an individual BZY grain boundary in a sample sintered at 2200°C [19] was experimentally observed with inline electron holography. EELS results indicate that barium vacancies may have contributed significantly to the potential well, while DFT calculations and GPA analysis suggest that GB lattice distortions also play an important role.

2) The negative sign of the potential is not in conflict with reported positive values of the potential in the literature for BZY grain boundaries. The literature values are based on impedance spectroscopy measurements and are only based on the potential due to redistribution of free charge carriers in the crystal. Inline holography is also sensitive to the mean inner potential of the crystal, which may yield a different sign.

3) EELS and EDX results suggest segregation of yttrium to the GB as a response to barium loss during fabrication. This however only slightly affects the total electrostatic potential.

4) The result suggests that the GB core is structurally distorted and has significant barium deficiency in a region around 2 nm wide, which may be linked to lower proton mobility across the GB.

5) If the goal is to measure the potential due to redistribution of free charge carriers, holography should be applied with caution: the large magnitude of the mean inner potential may mask variations in the potential due to redistribution of free charge carriers if both effects have a similar spatial extent.

## Acknowledgement

We would like to thank Phuong Nguyen and Adrian Lervik for TEM sample preparation. Computational resources have been provided by the Uninett Sigma2 National infrastructure for computational science in Norway. This work was supported by the project "Functional oxides for clean energy technologies" (FOXCET, 228355), and the Norwegian Centre for Transmission Electron Microscopy (NORTEM, 197405/F50) both funded by the Research Council of Norway.

# Appendix A

To find the bulk oxygen vacancy and proton concentrations inside the TEM column, consider the hydration reaction of BaZrO$_3$,

$$H_2O(g) + v_O^{\bullet\bullet} + O_O^x \leftrightarrow 2OH_O^{\bullet}, \quad (9)$$

with equilibrium constant

$$K_{\text{hydr}}(T) = \frac{[OH_O^{\bullet}]^2}{[v_O^{\bullet\bullet}][O_O^x]p_{H_2O}} = \exp[-(\Delta H_{\text{hydr}}^{\circ} - T\Delta S_{\text{hydr}}^{\circ})/k_B T], \quad (10)$$

where $p_{H_2O}$ is the water vapor pressure, $[v_O^{\bullet\bullet}]$ and $[OH_O^{\bullet}]$ the volume concentrations of oxygen vacancies and protons, and $\Delta H_{\text{hydr}}^{\circ}$ and $\Delta S_{\text{hydr}}^{\circ}$ the standard enthalpy and entropy of hydration. In this work, the thermodynamic parameters $\Delta H_{\text{hydr}}^{\circ} = -0.82$ eV and $\Delta S_{\text{hydr}}^{\circ} = -0.92$ meVK$^{-1}$ are used [1]. Three oxygen sites per formula unit yields the site restriction

$$[v_O^{\bullet\bullet}] + [OH_O^{\bullet}] + [O_O^x] = 3. \quad (11)$$

Furthermore, neglecting the contribution from electrons and holes, the charge neutrality condition becomes

$$2[v_O^{\bullet\bullet}] + [OH_O^{\bullet}] = [Y_{Zr}'], \quad (12)$$

where $[Y_{Zr}']$ is the volume concentration of yttrium substituents. Equations (10-12) are a set of three equations with three unknowns; its solution gives the oxygen vacancy and proton concentrations as functions of yttrium concentration, water vapor pressure and temperature. Figure 9 shows the nominal bulk concentrations of a 10 mol% Y-substituted BaZrO$_3$ as a function of $p$H$_2$O for $T$ = 300 K.

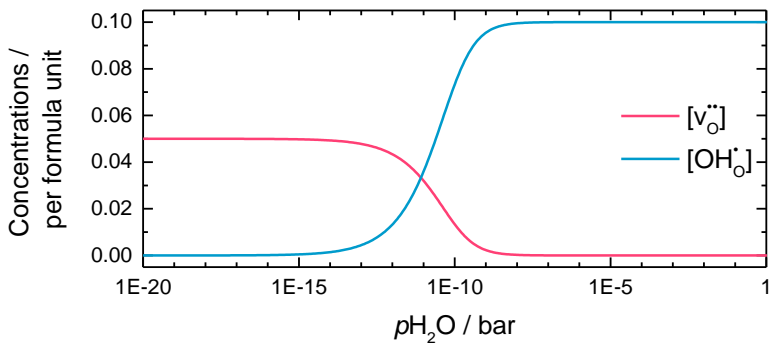

Figure 9. Nominal bulk concentrations of oxygen vacancies and protons in BaZr$_{0.9}$Y$_{0.1}$O$_3$ as a function of water vapor pressure, at $T$ = 300 K.

It is hard to estimate the exact $p$H$_2$O level in the TEM column, but it is clearly very low. The total pressure inside the column is $\sim 2 \cdot 10^{-10}$ bar, and the partial pressure of water should be lower for two reasons. First, assuming that the TEM column has the same relative water vapor content as the atmosphere at sea level gives $p\text{H}_2\text{O} \approx 7 \cdot 10^{-12}$ bar. Moreover, the cold trap in the TEM column inserted near the specimen stage enhances the vacuum even further. As the specimen stayed inside the TEM column for 12 hours before conducting the holography experiment, it is reasonable to assume the $v_O^{\bullet\bullet}$ and $\text{OH}_O^{\bullet}$ concentrations to have reached equilibrium. Hence, we can approximate the bulk defect concentrations to be $[v_O^{\bullet\bullet}] = 5$ mol% and $[\text{OH}_O^{\bullet}] = 0$ mol% inside the TEM column.

## Appendix B

Here, we show how Poisson's equation (3) is solved numerically to obtain $V_E$ in the space charge layers (SCLs). Assuming the GB to be homogenous and infinitely planar reduces the problem to one dimension, simplifying Poisson's equation. With *x* as the direction perpendicular to the GB plane, the space charge density can be taken as

$$\rho(x) = e(2[v_O^{\bullet\bullet}](x) + [\text{OH}_O^{\bullet}](x) - [Y_{Zr}']). \quad (13)$$

In the Mott-Schottky approximation, we assume a constant acceptor doping concentration in the SCL region, independent of *x*, which simplifies the further treatment considerably.

Using the oxygen site restriction in equation (11), and assuming constant electrochemical potential in the entire grain boundary region, the oxygen vacancy concentration in the SCL can be taken as

$$[v_O^{\bullet\bullet}](x) = \frac{3\,[v_O^{\bullet\bullet}] \exp\left[-\frac{2e\Delta V_E(x)}{k_B T}\right]}{3 + [v_O^{\bullet\bullet}]\left(\exp\left[-\frac{2e\Delta V_E(x)}{k_B T}\right] - 1\right) + [\text{OH}_O^{\bullet}]\left(\exp\left[-\frac{e\Delta V_E(x)}{k_B T}\right] - 1\right)}. \quad (14)$$

Similarly, the proton concentration in the SCL is

$$[\text{OH}_O^{\bullet}](x) = \frac{3\,[\text{OH}_O^{\bullet}] \exp\left[-\frac{e\Delta V_E(x)}{k_B T}\right]}{3 + [v_O^{\bullet\bullet}]\left(\exp\left[-\frac{2e\Delta V_E(x)}{k_B T}\right] - 1\right) + [\text{OH}_O^{\bullet}]\left(\exp\left[-\frac{e\Delta V_E(x)}{k_B T}\right] - 1\right)} \quad (15)$$

Note the dependence on $\Delta V_E(x)$: a larger potential will reduce the concentration of these positively charged defects.

A fundamental requirement is the material's overall charge neutrality, that is, that the sum of the SCL charge and the core charge is zero. From the above charge density, the total charge per area in the SCLs can be taken as

$$Q_{\text{SCL}} = 2 \cdot \int_0^\infty \rho(x)\,\mathrm{d}x \quad (16)$$

where $x = 0$ at the interface between the GB core and the SCL, and $\infty$ denotes a position in bulk, far away from the grain boundary. The factor 2 is included to account for both sides of the GB core. The per area charge at the GB core due to defect segregation is

$$Q_{\text{core}} = \sum_i z_i e n_i^{2D}, \tag{17}$$

where $z_i$ is the charge and $n_i^{2D}$ the two-dimensional concentration of defect *i*. To solve Poisson's equation, we apply the two boundary conditions $\Delta V_E'(\infty) = 0$ and $\Delta V_E(0) = \Delta V_{E,0}$. Initially, $\Delta V_{E,0}$ is guessed and adjusted with small increments until the charge neutrality condition $Q_{\text{SCL}} + Q_{\text{core}} = 0$ is satisfied. At this point we have obtained $V_E$ in the SCL.